\documentclass[9pt,twocolumn,twoside]{optica}
\setboolean{shortarticle}{false}
\setboolean{minireview}{false}
\usepackage{mathtools}

\title{Approaching the optimum phase measurement in the presence of amplifier noise}

\author[1,*]{Darko~Zibar}
\author[2]{Jens~E.~Pedersen}
\author[2]{Poul~Varming}
\author[1]{Giovanni Brajato}
\author[1]{Francesco Da Ros}

\affil[1]{DTU Fotonik, Technical University of Denmark, DK-2800, Kgs. ~Lyngby, Denmark}
\affil[2]{NKT Photonics, Blokken 84, DK-3460 Birkerød, Denmark}

\affil[*]{Corresponding author: dazi@fotonik.dtu.dk}

\begin{abstract} % 100 words

In fundamental papers from 1962, Heffener and Haus showed that it is not possible to construct a linear noiseless amplifier \cite{Heffener1950,Haus1962}. This implies that the amplifier intrinsic noise sources induce random perturbations on the phase of the incoming optical signal which translates into spectral broadening. Achieving the minimum induced phase fluctuation requires a phase measurement method that introduces minimum uncertainty i.e.~the optimum phase measurement. We demonstrate that a measurement method based on the heterodyne detection and the extended Kalman filtering approaches the optimum phase measurement in the presence of amplifier noise. A penalty of 5 dB (numerical) and 15 dB (experimental) compared to the quantum limited spectral broadening is achieved. Spectral broadening reduction of 44 dB is achieved, compared to when using the widely employed phase measurement method, based purely on the argument of the signal field. Our results reveal new scientific insights by demonstrating a phase measurement method that enables to approach the minimum phase fluctuation, induced by the amplifier noise. An impact is envisioned for phase-based optical sensing systems, as optical amplification could increase sensing distance with minimum impact on the phase.

\end{abstract}

\setboolean{displaycopyright}{true}

\begin{document}
\maketitle
%%%%%%%%%%%%%%%%%%%%%%%%%%%%%%%%%%%%%%%%%%%%%%%%%%%%%%%%%% section 1 %%%%%%%%%%%%%%%%%%%%%%%%%%%%%%%%%%%%%%%%%%%%%%%%%%%%%%%%%%
\section{Introduction} \label{sec:intro}
High-power narrow-linewidth lasers are essential for optical fibre sensing, gravitational wave detection, optical space communication and high-capacity optical fibre communication \cite{Sun:20,Droste,Bode:20,Zhang:20,Wellmann:19,Trichili:20,Wellmann:21}. Typically, a high-power performance is achieved by amplifying a low-noise seed laser with a single or several stages of fiber-optic amplification. 

Apart from increasing the amplitude, an optical amplifier will induce random fluctuations on the phase of the incoming optical signal. The minimum induced phase fluctuation, (quantum limit), due to the amplifier noise, denoted by $\Delta \phi_{ql}$, is a consequence of Heisenberg's uncertainty principle \cite{Heffener1950,Haus1962}. Moreover, to reach the minimum value, the detector must measure photon number and phase with the same relative uncertainties as those introduced by the amplifier, i.e.~optimum detector \cite{Heffener1950,Haus1962}. As the optical amplifier reduces the signal-to-noise-ratio of the incoming signal, the magnitude of $\Delta \phi_{ql}$ is inversely proportional to it. The random fluctuation of the optical phase, $\Delta \phi_{ql}$, will translate into spectral broadening denoted by $\Delta \nu_{ql}$.  For the applications of high-power narrow-linewidth lasers in optical sensing and communication increasing the laser's spectral width translates into system performance degradation, and it should therefore be kept at its minimum. 

In order to quantify the impact of the amplifier noise, a measurement of the optical phase and the corresponding spectral width, with and without the optical amplification, and the comparison to the quantum limits $\Delta \phi_{ql}$ and $\Delta \nu_{ql}$ is needed. A practical optimum phase measurement method is needed to approach $\Delta \phi_{ql}$ and $\Delta \nu_{ql}$ and thereby minimize the impact of the amplifier noise.  

So far, the approach for quantifying the impact of amplifier noise has been based on the measurement of the phase power spectrum density (PSD) before and after the amplification \cite{Xue2020,Moller1998,Wellmann:21,Rochat2001,soton77479,desurvire2002erbium,Ricciardi:13}. However, the reported experimental results do not fully agree, and are even contradictory, in some cases. What makes the measurements challenging is that the impact of the amplifier noise is mostly visible at high frequencies (>10MHz) of the phase PSD and for low input signal power levels. The state-of-the-art optical phase measurement methods, including spectrum analyzers, suffer from the limited sensitivity as well as the limited dynamic and frequency range. This makes it challenging to measure optical phase below the thermal measurement noise floor, which typically lies around -150 dB rad$^2$/Hz, and at frequencies exceeding a few MHz. Thus, observing and quantifying the impact of amplifier noise requires a highly-sensitive optical phase measurement method. Finally, to fully quantify the impact of the amplifier noise a comparison to the quantum limit $\Delta \phi_{ql}$ is necessary.

In this paper, we investigate the impact of the amplifier noise on the phase fluctuation and the spectral broadening of the incoming optical signal, as a function of the input signal power into the amplifier, and compare it to the quantum limits. For measuring the optical phase, a heterodyne receiver, in a combination with the extended Kalman filter, is proposed. The phase measurement method is a practical implementation of the theoretically optimal phase measurement in the presence of the amplifier noise. We demonstrate a highly accurate phase measurement resulting in a penalty of 5 dB (numerically) and 15 dB (experimentally) compared to the quantum limit. Finally, we show that the commonly employed phase measurement method results in a large penalty for the increasing system bandwidth and decreasing signal power.

%for a wide range of input signal power levels and bandwidths, numerically and experimentally, an optimum heterodyne phase measurement method that would result in a minimum impact of the amplifier noise. The proposed method operates closely to the theoretical lower limit. We also show numerically and experimentally that if the same phase estimation as in \cite{Heffener1950} is employed the Heffner limit is indeed reached, but it will not always be a lower limit. To the best of our knowledge, this is the first experimental demonstration of the Heffner limit.    
The outline of the paper is as follows: In section \ref{sec:theory}, a theoretically achievable minimum (quantum limited) variance for the phase fluctuations, due the amplifier noise, is presented for the conditions under the consideration. We also present and provide practical method, based on the extended Kalman filtering, for approaching the optimum phase estimation and achieving the minimum phase fluctuation. In section \ref{sec:numerical}, numerical results are presented and compared to the quantum limits. We also illustrate the impact of the amplifier noise by computing the phase PSDs. In section \ref{sec:exp_results}, the impact of amplifier noise on the phase PSD is experimentally investigated. Finally, the corresponding spectral broadening is computed and compared to the quantum limit. Finally, in section \ref{sec:conc}, the main findings are summarized and the impact of results is discussed.       
%%%%%%%%%%%%%%%%%%%%%%%%%%%%%%%%%%%%%%%%%%%%%%%%%%%%%%%%%% section 2 %%%%%%%%%%%%%%%%%%%%%%%%%%%%%%%%%%%%%%%%%%%%%%%%%%%%%%%%%%
\section{Theoretical framework}\label{sec:theory}
\subsection{Minimum phase fluctuation of an amplified time-varying phase signal}
In the analysis to follow, only intrinsic (fundamental) noise sources are included. This assumption is valid as we are interested in the minimum phase fluctuation induced by the amplifier noise.

We assume that the input to the linear amplifier is provided by a laser sources with the angular frequency $\omega = 2\pi\nu$ and a randomly time-varying phase $\phi(t)$:

\begin{equation}\label{eq:input sig}
    E_{out}(t) =\sqrt{P_{in}}\sin[\omega t + \phi(t)] 
\end{equation}

where $P_{in}$ is the input signal power. The minimum fluctuation of $\phi(t)$ is governed by the amplified spontaneous emission within the laser cavity (quantum noise) and can be approximated by a Wiener process \cite{coldren2012diode}:

\begin{equation}\label{eq:Wiener process}
    \frac{d \phi}{dt} = \sqrt{\kappa}\Gamma(t)
\end{equation}

where $\kappa$ is the rate of phase diffusion and is related to the laser coherence time $\tau_c$. $\Gamma(t)$ is a Langevin noise source with Gaussian distribution. It has a zero mean and a delta correlation function. The PSD of a laser signal with a time-varying phase described by a Wiener process has a Lorentzian shape and is centered around $\omega$. Its spectral width, $\Delta \nu_c$, is inversely proportional to the coherence time, $\tau_c$ and can be computed using:

\begin{eqnarray}\label{eq:spectwidth}
    \Delta \nu_c &=& \frac{\bigg(\int^{\infty}_{0}S(\nu) dv\bigg)^2}{\bigg(\int^{\infty}_{0}S^2(\nu) dv\bigg)} \\ \nonumber 
    &=& \Bigg[\int_{-\infty}^\infty \bigg|\frac{\langle U^\ast(t)(t)U(t)(t+\tau)\rangle}{\langle U^\ast(t)U(t)\rangle}\bigg|^2d\tau\Bigg]^{-1}=\frac{1}{\tau_c}
\end{eqnarray}

where $S(\nu)$ is the PSD of the laser signal $E_{in}(t)$ and $U(t)=\sqrt{P_{in}}e^{j(\omega t+\phi(t))}$.

%The incoming signal $E_{in}$ is amplified with a linear amplifier with a gain $G$ and the minimum noise power $P_N= h\nu(G-1)B$, where $B$ is the bandwidth. The output of the amplifier is expressed as a coherent superposition of the amplified signal and the amplifier noise:

%\begin{equation}\label{eq:sig+noise}
%    E_{out}(t) =\sqrt{P_s}\sin[\omega_0 t + \phi_0(t)] + n_a(t) 
%\end{equation}

%where $P_s= GP_{in}$ and $n_a(t)$ is a noise term, added by the amplifier, with a zero mean Gaussian distribution and variance (noise power) $P_N$. 

The signal $E_{in}(t)$ is now applied to the linear amplifier with a gain, $G$. The output of the amplifier can be expressed as a coherent superposition of the amplified signal and the amplifier noise \cite{Heffener1950,Haus1962}:

\begin{equation}\label{eq:sig+noise}
    E_{out}(t) =\sqrt{P_{s}}\sin[\omega t + \phi(t)] + n_a(t) 
\end{equation}

where $P_s= GP_{in}$. $n_a(t)$ is a noise term, added by the amplifier. It has a zero mean Gaussian distribution and a minimum noise power $P_N= h\nu(G-1)B$ \cite{Heffener1950,Haus1962}. $B$ is the bandwidth and $h$ is Planck's constant.

To quantify the impact of the amplifier noise on the phase of the incoming signal, $\phi(t)$, a receiver with bandwidth $B$ is used. The receiver must employ a phase estimation method that introduces minimum uncertainty in the presence of the amplifier noise. This is accomplished by performing a statistically optimum, (and thereby theoretically the most accurate), phase estimation. It is achieved by solving the following Maximum a Posteriori (MAP) phase estimation problem \cite{Sarka}: 

\begin{eqnarray}\label{eq:MAP problem}
    \phi^{MAP}(t) &=& \arg \max_{\phi'} p[\phi'(t)|E_{out}(t)] \\ \nonumber
    &=& \arg \max_{\phi'} \frac{p[E_{out}(t)|\phi'(t)]p[\phi'(t)]}{p[E_{out}(t)]}
\end{eqnarray}

where $\phi'(t)$ represent various phase evolutions (different trail phases). $p(\cdot)$ and $p(\cdot|\cdot)$ denote the marginal and the conditional probability densities, respectively. Analytical solution to Eq.~(\ref{eq:MAP problem}) does not exist, due to the nonlinear relationship between $E_{out}(t)$ and $\phi(t)$ (nonlinear estimation problem). To find the analytical solution, linearization needs to be performed.

Assuming that the phase diffusion constant $\kappa$ is small enough such that $\sin[\phi(t)] \approx \phi(t)$, and setting $\omega=0$ for the convenience, estimating the phase becomes a linear estimation problem since $E_{out}(t) \propto \phi(t) + n_a(t)$ which has an analytical solution \cite{Haykin:2002}.  Under the aforementioned approximation, the solution to Eq.~(\ref{eq:MAP problem}) is obtained by finding a phase $\phi'(t)$ that minimizes the mean square error (MSE) defined as $E[(\phi(t) - \phi'(t))^2]$. The phase that is the solution to the MSE is denoted as: $\phi'(t)=\phi^{MAP}(t)$. The minimum phase fluctuation due to the amplifier noise is then expressed as:

\begin{equation}
    \Delta \phi^{MAP}_a = \sqrt{E[(\phi(t) - \phi^{MAP}(t))^2]}
\end{equation}

The results from the quantum phase estimation state that the expression for the minimum phase fluctuation, $\Delta \phi^{MAP}_a$, then becomes \cite{Berry,Tsang}:

\begin{equation} \label{eq:min phase MAP}
    \Delta \phi^{MAP}_a = \frac{1}{\sqrt{\sqrt{2N_p}}}
\end{equation}

where $N_p$ describes the mean number of photons per laser coherence time and is defined as:

\begin{equation}
    N_p = \frac{P_s\tau_c}{h\nu(G-1)} = \frac{P_s}{h\nu(G-1)\pi \Delta \nu_{FWHM}}
\end{equation}

where $\tau_c= 1/\pi \Delta \nu_{FWHM}$ is a coherence time of a laser with phase modelled as a Wiener process and  $\Delta \nu_{FWHM}$ is the full width at half maximum (FWHM) of the Lorentzian laser spectrum denoted by $S(\nu)$. It should be stressed that the minimum phase fluctuation expressed by Eq.~(\ref{eq:min phase MAP}) assumes that the receiver bandwidth $2B$ is matched to the laser spectral width $\Delta \nu_c = \Delta \pi\nu_{FWHM}$ \cite{Berry,Tsang}.

We now need to find a practical receiver architecture that can perform statistically optimal phase estimation and approach $\Delta \phi^{MAP}_a$. In the following, we consider two different cases.

\subsection{Minimum phase fluctuation of an amplified single frequency constant phase signal}
If we assume a constant phase signal ($\phi(t)=\phi$), the PSD becomes a delta function centered around $\omega$. Applying a small detector bandwidth, $B$, such that the power spectrum density of the amplified signal is confined to a relatively narrow band in the neighborhood the signal frequency (high SNR condition), the optimum phase estimation becomes \cite{Rice,Henry}:

\begin{eqnarray} \label{eq:tan_phase_est}
\phi_{\tan^{-1}} &=& \arg[(E_{out}(t) + j \mathcal{H}\{E_{out}(t)\})e^{-j\omega t}] \nonumber \\
&=&\arg[(I +jQ)e^{-j\omega t}] = \tan^{-1}(Q/I)
\end{eqnarray}

where $\mathcal{H}\{\cdot\}$ denotes Hilbert transform. Using Eq.~(\ref{eq:tan_phase_est}), the resulting minimum phase fluctuation (standard deviation) due to the amplifier noise can then be computed to as \cite{Rice,Heffener1950}:

\begin{equation}\label{eq:phi_a}
\Delta \phi_a = \sqrt{\frac{P_N}{2P_s}}=\sqrt{\frac{h\nu(G-1)B}{2P_s}} = \sqrt{\frac{1}{2SNR}} = \frac{1}{\sqrt{2N_p}}
\end{equation}

where $N_p = \frac{P_s}{h\nu(G-1)B}$. The phase estimation method in Eq.~(\ref{eq:tan_phase_est}) and the subsequent limit in Eq.~(\ref{eq:phi_a}) is widely employed in the literature for investigating the impact of amplifier noise \cite{Xue2020,Moller1998,Wellmann:21,Rochat2001,soton77479,desurvire2002erbium,Ricciardi:13}. However, for many practical applications within optical metrology and transmission, a relatively large receiver bandwidth is needed (up to a few GHz) and the input signal power to the amplifier may be low.\footnote{Generally speaking, the broader the receiver bandwidth, the larger is the number of photons, associated with the amplifier noise, with random phases that combine with the laser mode. This can lead to a large spectral broadening unless we find an effective receiver able to distinguishing between the photons associated with the laser phase $\phi(t)$ and the amplifier noise.} Finally, the constant phase condition is hard to satisfy in practice as the input to the amplifier is always provided by a laser source that has phase noise (randomly time-varying phase). Under the aforementioned practical conditions the phase estimation provided in Eq.(\ref{eq:tan_phase_est}) is no longer optimal and the corresponding minimum phase fluctuation governed by Eq.(\ref{eq:phi_a}) does no longer present a lower bound for the minimum phase fluctuation.

\subsection{Bayesian filtering for optimum phase estimation}
For the practical case of large signal bandwidth, low signal powers and time-varying phase, the state-spaced based Bayesian filtering is a highly-accurate and a practical approach for the MAP estimation problem in Eq.~(\ref{eq:MAP problem}). The state-space based Bayesian filtering implements a recursive MSE approach for estimating the quantity of interest and can be easily implemented in analog or digital domain \cite{Sarka}. To perform the estimation, the state-space model needs to be defined. It consists of a measurement and a state equation. The measurement equation describes the relationship between the observable (measurable) quantity and the parameter we want to estimate, i.e.~Eq.~(\ref{eq:sig+noise}). the state equation describes the evolution of the quantity to be estimated, in our case the phase, i.e.~Eq.(\ref{eq:Wiener process}). Excellent treatment of Bayesian filtering techniques is provided in Reference \cite{Sarka}.     

In this paper, we consider shot-noise limited heterodyne coherent detection in combination with digital signal processing for phase estimation. This implies that the amplified optical signal is combined with an (ideal) local oscillator (LO) laser ($\kappa=0$) in a 3-dB coupler, detected by the shot-noise limited balanced receiver and sampled by an analogue-to-digital converter. The state-space model, including the state and the measurement equation, Eq.~(\ref{eq:ssm state}), and (\ref{eq:ssm measure}) for the phase estimation, is then expressed as:

\begin{eqnarray}
\phi[k] & = & \phi[k-1] + \Gamma[k-1] \label{eq:ssm state}\\
y[k] & = & 2R\sqrt{P_sP_{LO}}\cos(\Delta \omega k T_s + \phi[k])  + n^{sh}[k] + n^b[k]\label{eq:ssm measure}
\end{eqnarray}

where $k=1,...,K$ is an integer representing the discrete--time, $P_{LO}$ is the LO power. $R$ is the responsivity of photodetector assumed to be 1. $\Delta \omega$ is the difference angular frequency between the signal and the LO laser, $T_s$ is the sampling time, $\phi[k]$ is the laser source time--varying phase modelled as a Wiener process with a phase diffusion constant $\kappa^2=2 \pi \Delta \nu_{FWHM} T_s$. The shot and the beat noise terms, $n_{sh}[k]$ and $n_{b}[k]$, are modelled as Gaussian noise sources with zero mean and variances: $\sigma^2_{shot}=2qR(P_{LO})B$ and $\sigma^2_{b}=4P_{LO}h\nu(G-1)B$, respectively \cite{Essiambre:10}. $q$ is the elementary charge and $B$ is the 3-dB bandwidth of the balanced receiver. The state equation, (\ref{eq:ssm state}), represents the discrete time Wiener process. \vspace{0.15cm}  

Exact solution to the phase estimation problem, given the state-space model described by Eq.~(\ref{eq:ssm state}) and (\ref{eq:ssm measure}) is not achievable. The reason is the nonlinear relationship between $y_k$ and $\phi_k$. Therefore, approximation to the Bayesian filtering in terms of extended, cubature or unscented Kalman filter needs to be employed \cite{Sarka}. In this paper, we choose to implement the extended Klaman filter (EKF) due to its faster processing time compared to the unscented and cubature Kalman filters. The state-space model described by Eq.(\ref{eq:ssm state}) and (\ref{eq:ssm measure}) can be directly plugged in into EKF filtering equation Eq.~(5.26)-(5.27) in \cite{Sarka}. The EKF will explore correlation properties of $\phi[k]$ specified by the state equation for tracking and estimation purposes, while ignoring the contribution from the amplifier noise since there is no correlation between $\phi[k]$ and $n_a[k]$ and $n_b[k]$.    

The estimated phase by the EKF will be denoted as $\phi_{EKF}[k]$. For a comparison reasons, we also apply the phase estimation method expressed by Eq.~(\ref{eq:tan_phase_est}), and denote the estimated phase by $\phi_{\tan^{-1}}[k]$. Using the aforementioned phase estimators, the phase fluctuation due to the amplifier noise is then computed as a sample standard deviation:

\begin{equation}\label{eq:sample std}
    \sigma_{EKF/\tan^{-1}} = \sqrt{\frac{1}{K}\sum_{k=1}^K(\phi[k]-\phi_{EKF/\tan^{-1}}[k])^2}
\end{equation}

\begin{figure}[t]
\centering
{\includegraphics[width=1\linewidth]{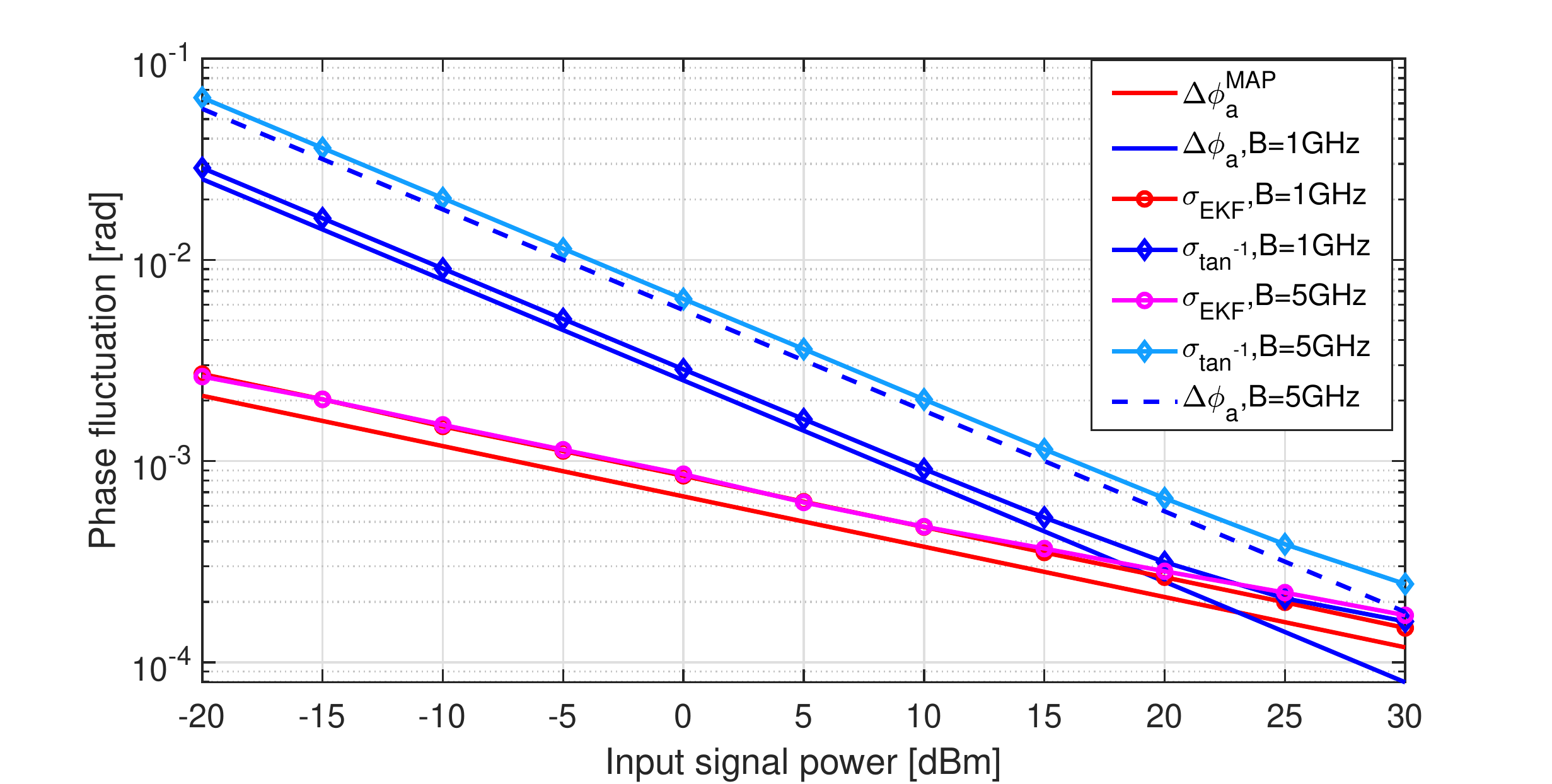}}
\caption{\textit{Simulation:} Phase fluctuation as a function of input signal power $P_s$.}
\label{fig:MSEvsPin}
\end{figure}

\begin{figure}[t]
\centering
{\includegraphics[width=1\linewidth]{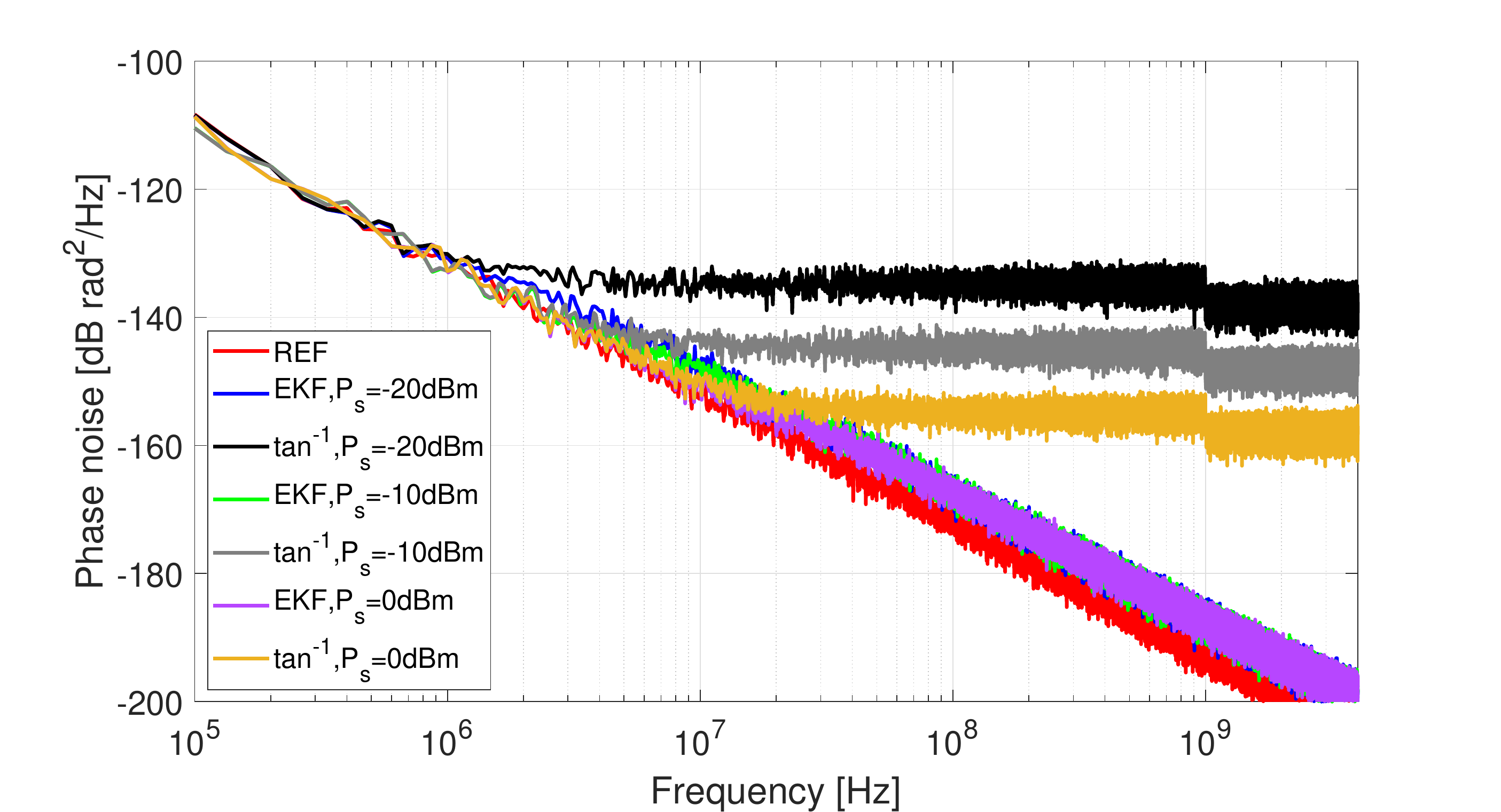}}
\caption{\textit{Simulation:} Phase power spectral density as a function of frequency using different phase estimation methods.}
\label{fig:FNvsPin}
\end{figure}

\subsection{Translating phase fluctuation into spectral broadening}
Given an observation time $T=kT_s$ under which the time-varying phase is observed and a laser source with Lorentzian spectrum, the phase fluctuation, $\Delta \phi$, results in the following spectral broadening, $\Delta \nu$ \cite{desurvire2002erbium}:

\begin{equation}\label{eq:spect broadening}
    \Delta \phi = \sqrt{\frac{2\pi \Delta \nu_c T}{\pi}}  \Longrightarrow \Delta \nu_c = \frac{\Delta \phi^{2}}{2 T}
\end{equation}

\section{Numerical results}\label{sec:numerical}
In Fig.~\ref{fig:MSEvsPin}, the minimum theoretically achievable (quantum noise limited) phase fluctuation due to the amplifier noise, $\Delta \phi^{MAP}_a$, and $\Delta \phi_a$, is plotted. We also plot the numerically computed phase fluctuation using Eq.~(\ref{eq:sample std}), for the phase estimation based on the EKF and Eq.~(\ref{eq:tan_phase_est}), i.e. $\sigma_{EKF}$ and $\sigma_{\tan^{-1}}$, respectively. We consider the receiver bandwidths $B$ of 1 GHz and 5 GHz. The laser linewidth $\Delta \nu_{FWHM}$ is chosen to be 10 Hz. The reason for choosing relatively small linewidth is that we would like to minimize the penalty to $\Delta \phi^{MAP}_a$, which is derived for small diffusion factors (small linewidths).    

It should be noticed from Fig.~\ref{fig:MSEvsPin} that the minimum phase fluctuation $\Delta \phi^{MAP}_a$ is significantly lower compared to $\Delta \phi_a$ for $B=5$ GHz. As the bandwidth is reduced to $B=1$ GHz, $\Delta \phi^{MAP}_a$ still provides a lower phase fluctuation for the input signal power levels up to 20 dBm.  

The numerically computed phase fluctuation $\sigma_{EKF}$, exhibits only 5 dB of penalty compared to the theoretically achievable limit given by $\Delta \phi^{MAP}_a$. Most importantly, it does not increases as the bandwidth is increased from 1 GHz to 5 GHz. Part of the penalty is due to the variance of the beat term, $n_b[k]$, being $4P_{LO}$ times larger than the variance of $n_a[k]$. In general, the results for $ \sigma_{EKF}$ illustrate the robustness of the EKF method for estimating the phase in the presence of the amplifier noise and receiver bandwidths significantly larger than the spectral width. Finally, the numerically computed phase fluctuation $\sigma_{\tan^{-1}}$ closely follows the theoretical limit given by $\Delta \phi_a$. The penalty of approximately 1 dB, compared to $\Delta \phi_a$, is due to the variance of the beat term. Also, it should be noted that both $\Delta \phi_a$ and $\sigma_{\tan^{-1}}$ increase as the bandwidth is increased.

From the practical point of view, quantifying the impact of the amplifier noise in terms of the phase fluctuation is very challenging. A more practical approach would be to investigate the impact of the amplifier noise on the phase PSD, and then compute the signal spectral broadening and compare it to the quantum limit. 

In Fig.~\ref{fig:FNvsPin}, the phase PSD obtained using the phase estimation method based on the EKF and Eq.~(\ref{eq:tan_phase_est}) is plotted as a function of frequency. We also plot a reference PSD computed using $\phi[k]$. The input signal power, $P_s$, is varied from -20 dBm to -5 dBm. Using the phase estimation based on Eq.~(\ref{eq:tan_phase_est}), the impact of the amplifier noise is clearly observable. The amplifier noise results in a horizontal noise floor which increases as $P_s$ is decreased. 

The phase PSD computed using the EKF based phase estimation method resembles, to a high degree, the reference spectra. The impact of the amplifier noise is visible as the deviation from the reference spectrum is observed. However, the deviation is not nearly as large as when Eq.~(\ref{eq:tan_phase_est}) is employed for the phase estimation. This illustrates the efficiency of the phase estimation method based on the EKF. 

\begin{figure}[t]\label{fig:spec broad vs Pin}
\centering
{\includegraphics[width=1\linewidth]{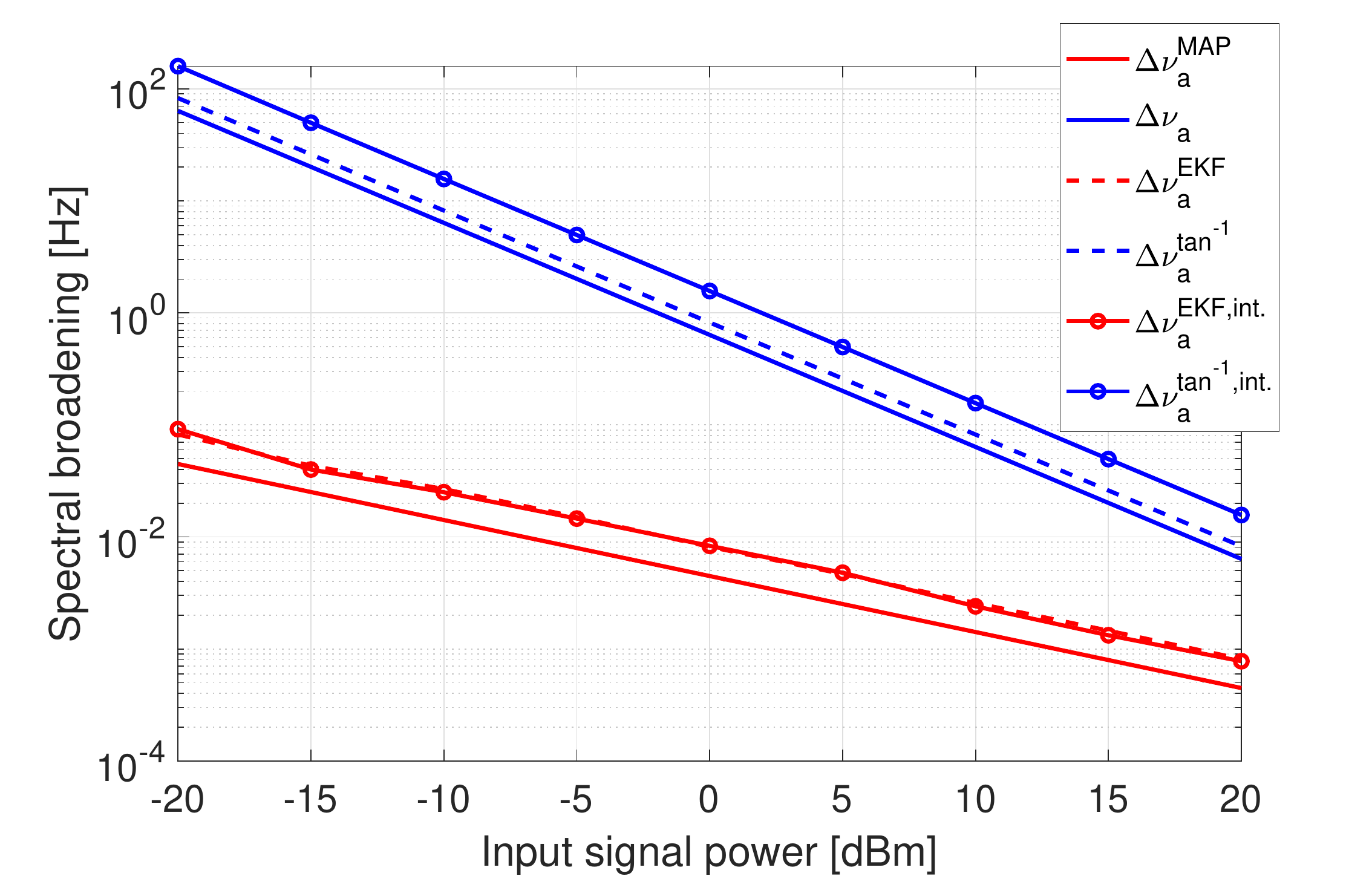}}
\caption{\textit{Simulation:} Spectral broadening as a function of input signal power, $P_s$, for different phase estimation methods.}
\end{figure}

\begin{figure*}[t]
  \centering
  \includegraphics[width=1\textwidth]{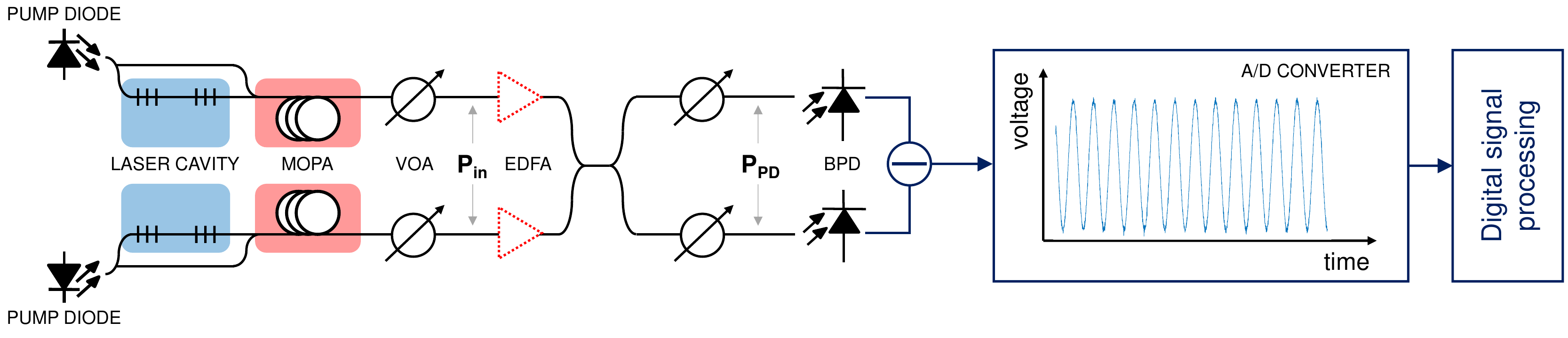}
\caption{Experimental set--up for investigating the impact of amplifier noise on the phase fluctuation of the incoming signal.}
\label{fig:exp setup}
\end{figure*}

\begin{figure*}[t]
  \centering
  \includegraphics[width=1\textwidth]{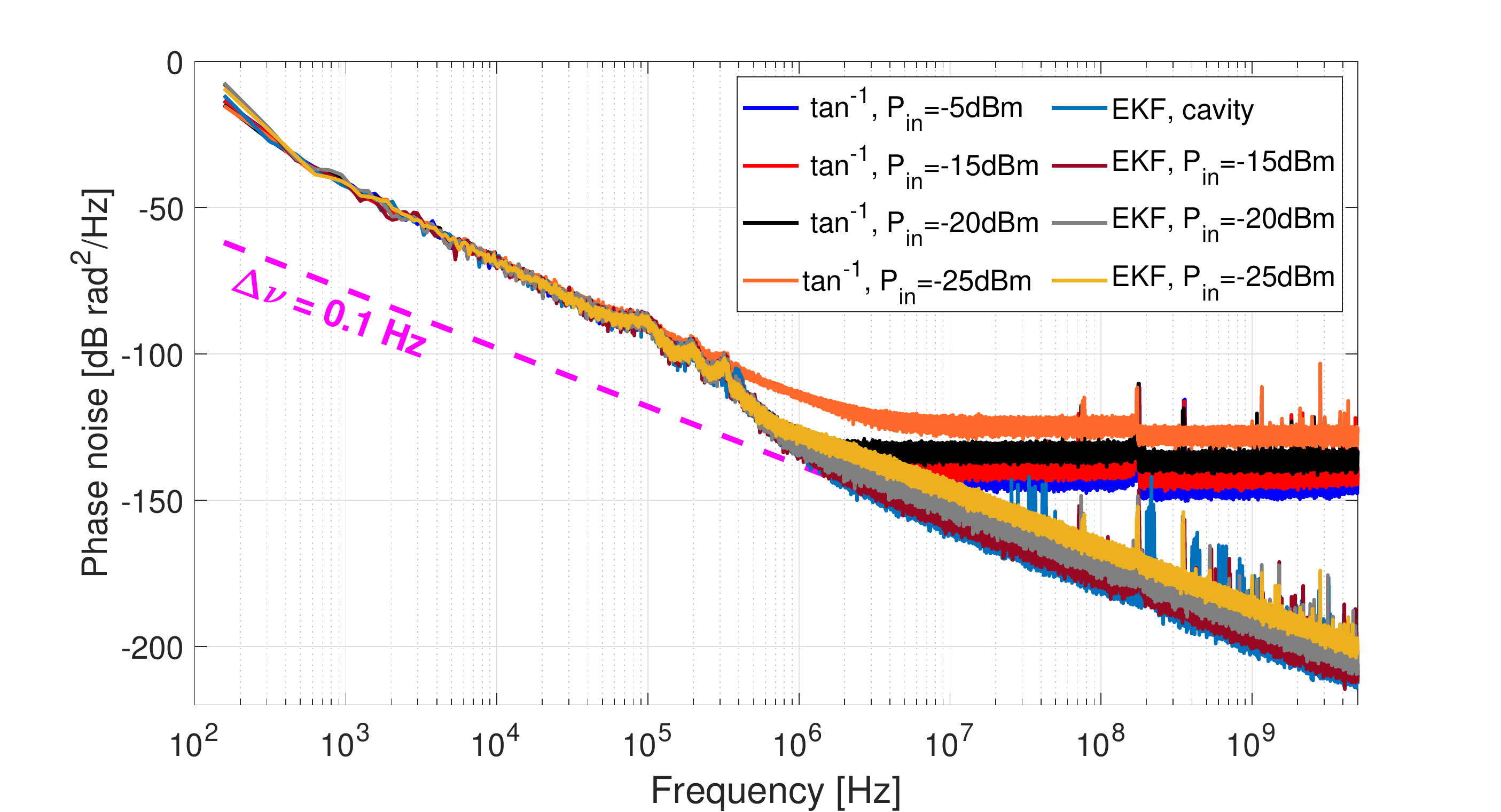}
\caption{\textit{Experiments:} Phase power spectral density as a function of frequency using different phase estimation methods and input power to the EDFA.}
\label{fig:FN spectra exp}
\end{figure*}

\begin{figure}[t]
\centering
{\includegraphics[width=1\linewidth]{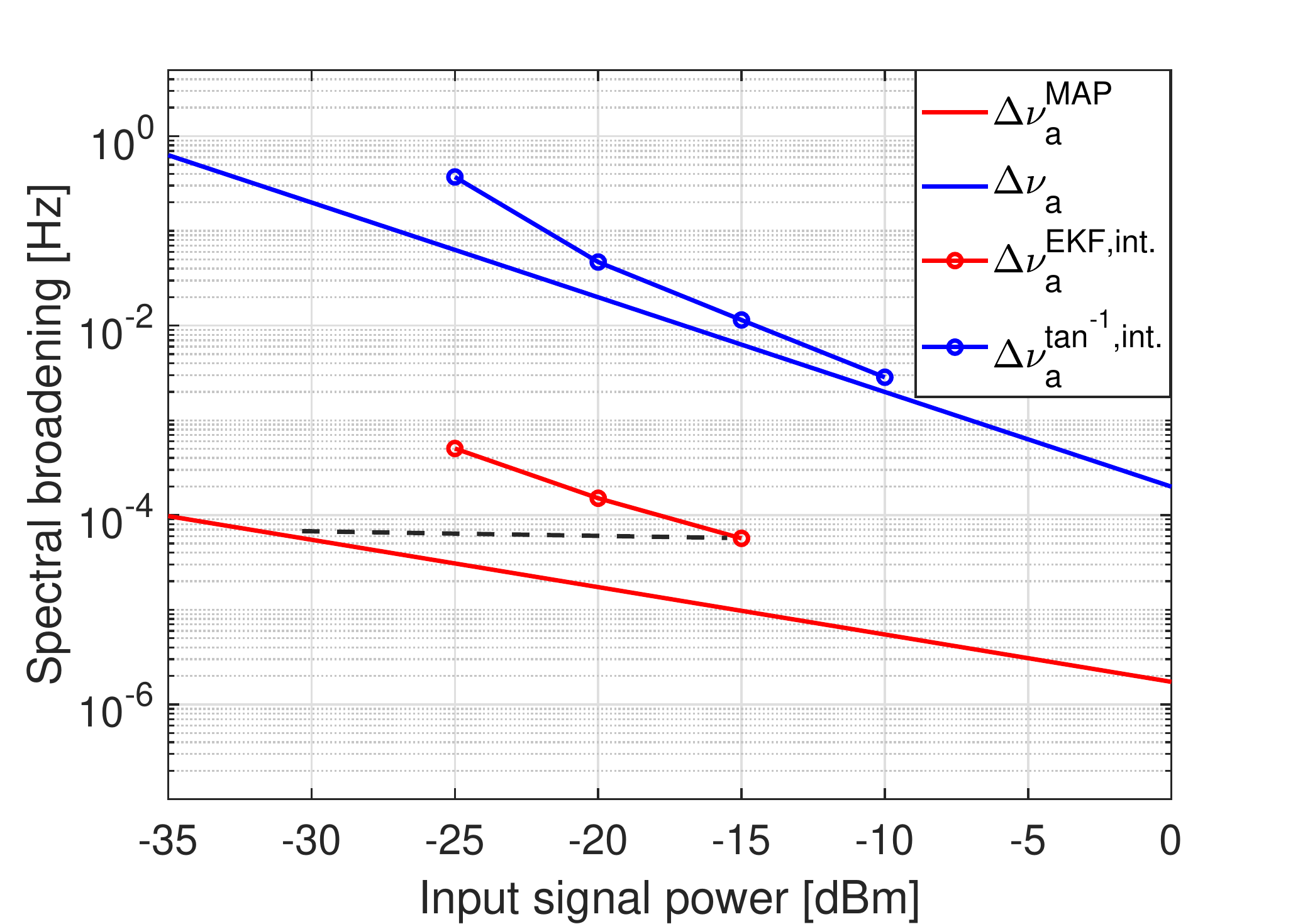}}
\caption{\textit{Experimental:} Spectral broadening as a function of input signal power, $P_{in}$, for different phase estimation methods.}
\label{fig:spec broad vs Pin exp}
\end{figure}

Next, we compute the corresponding spectral broadening, due to the amplifier noise as a function of the input signal power $P_s$. This is achieved by performing the numerical integration of phase PSDs. The corresponding spectral widths are denoted by $\Delta \nu^{EKF,int.}_a$ and $\Delta \nu^{\tan^{-1},int.}_a$, respectively. The theoretical values for the minimum spectral broadening, (quantum limit), denoted by $\Delta \nu^{MAP}_a$ and $\Delta \nu_a$ are obtained by inserting Eq.~(\ref{eq:min phase MAP}) and (\ref{eq:phi_a}) in Eq.~(\ref{eq:spect broadening}), respectively.

%The theoretical values are obtained by inserting Eq.~(\ref{eq:phi_a}) and (\ref{eq:min phase MAP}) in (\ref{eq:spect broadening}) and are denoted by $\Delta \nu_a$ and $\Delta \nu^{MAP}_a$, respectively. Given the estimated phase $\phi_{EKF}[k]$ and $\phi_{\tan^{-1}}[k]$, we first compute the corresponding spectral width using the numerical integration of the autocorrelation function in Eq.~(\ref{eq:spectwidth}). The obtained values are then subtracted from the reference spectral width, $\Delta \nu_0$, computed for the phase $\phi_0[k]$. 

To investigate the accuracy of the numerical integration, we convert the numerically computed phase fluctuations, $\sigma_{EKF}$, and, $ \sigma_{\tan^{-1}}$, into spectral broadening using Eq.~(\ref{eq:spect broadening}). The corresponding spectral broadening are denoted by $\Delta \nu^{EKF}_a$ and $\Delta \nu^{\tan^{-1}}_a$, respectively. The agreement between $\Delta \nu^{EKF,int.}_a$ and $\Delta \nu^{EKF}_a$ indicates that the error induced by the numerical integration is small. A penalty of approximately 3 dB is observed between $\Delta \nu^{\tan^{-1},int.}_a$ and $\Delta \nu^{\tan^{-1}}_a$. A possible reason is that the $\phi_{\tan^{-1}}[k]$ contains a higher degree of noise compared to $\Delta \nu^{EKF}_a$, which makes the numerical differentiation becomes less accurate. 

It should be emphasized that the spectral broadening $\Delta \nu^{EKF,int.}_a$ is significantly below $\Delta \nu_a$ and $\Delta \nu^{\tan^{-1}}_a$. Finally, a penalty of only 5 dB is observed between $\Delta \nu^{EKF,int.}_a$ and a minimum spectral broadening $\Delta \nu^{MAP}_a$. 

%%%%%%%%%%%%%%%%%%%%%%%%%%%%%%%%%%%%%%%%%%%%%%%%%%%%%%%%%% section 2 %%%%%%%%%%%%%%%%%%%%%%%%%%%%%%%%%%%%%%%%%%%%%%%%%%%%%%%%%%

\section{Experimental results} \label{sec:exp_results} 
The experimental set-up for investigating the impact of the amplifier noise on the phase fluctuation and the corresponding spectral broadening is shown in Fig.~\ref{fig:exp setup}. 

The state-space model, employed for the extended Kalman filtering based phase estimation, is described by Eq.~(\ref{eq:ssm state}) and (\ref{eq:ssm measure}). The state equation (\ref{eq:ssm state}) assumes that the laser is only dominated by the quantum noise which is typically not the case for practical laser sources. Practical laser source will not only be dominated by the intrinsic noise sources but also by technical/environmental noise. The impact of these is that the penalty may be expected.

We employ two similar fiber lasers, one as a signal, and the other as the LO source. Both lasers have matching phase noise performances. The lasers use a fiber Bragg grating cavity to produce an output beam operating in a single mode. The maximum output power measured directly from the cavity is -2 dBm. The laser cavity is then followed by a master oscillator fiber amplifier (MOPA) to boost the signal power level up to 15 dBm. The signal and LO lasers are set at 1550.08~nm, with a frequency offset of approx.~200~MHz, in order to minimize the impact of the DC response of the electronics at the receiver. 

The lasers modules can output the signal directly after the cavity and the MOPA, see Fig.~\ref{fig:exp setup}. The output signals right after the cavity give us the opportunity to measure a reference phase PSD and its corresponding spectral width. Finally, after the signal and the LO laser modules (cavity plus MOPA), we add an additional Erbium Doped fiber amplifier (EDFA). The EDFAs are commercial dual-stage low-noise pre-amplifiers (noise figure approx.~5~dB) operated at maximum pump currents, corresponding to an target output power of approx.~15~dBm. Variable optical attenuators placed before the EDFAs allow to tune the input power $P_{in}$ and thereby to vary the SNR to the receiver.

The two optical carrier are then combined in a 50:50 coupler and injected into the two inputs of a 43-GHz balanced photodetector used for coherent signal detection. A second set of VOAs at each photodetector inputs allow to keep the input power into each photodetector arm constant to $P_{PD} = 0.64$~mW.

%\textbf{@ DARKO: I also have all the values of laser cavity power output/ laser+MOPA output/ laser RIN settings etc. We can have a chat about which values are meaningful to mention. Maybe some of these should be added to the Supplementary material.}

The photodetector is followed by a digital storage oscilloscope (DSO) performing the analog-to-digital (A/D) conversion. The DSO has a bandwidth of 13~GHZ and operates at a sampling rate of $F_s = 40 GHz$. The sampled signal is then stored for offline signal processing. The memory of the sampling scope allows us to store $K=256\times10^{6}$ samples. This results in the minimum phase PSD frequency of $f_{min} = F_s/K = 156$~Hz, and a maximum PSD frequency of $f_{max} = F_s/2 = 20$~GHz. Similarly to the numerical results, we use the EKF and Eq.~(\ref{eq:tan_phase_est})($\tan^{-1}$)--based phase estimation methods. 

In Fig.~\ref{fig:FN spectra exp}, the phase PSD is shown as a function of frequency. To obtain the reference phase PSD, the laser outputs (signal and LO) right after the cavities are measured. We use the EKF method for the phase measurement due to its higher accuracy. Assuming equal contributions from the signal and the LO laser, the obtained phase PSD is divided by a factor of 2. The reference phase PSD illustrates that the quantum limited laser linewidth (Schawlow-Townes linewidth) is around 0.1 Hz. 

Next, the input signal and the LO power to the EDFA amplifiers is decreased from -20 dBm to -15 dBm while keeping the signal power to the balanced receiver at a constant level. A clear impact of the amplifier noise on the phase fluctuation and thereby the corresponding phase PSD is observed for $\tan^{-1}$ phase estimation method. Moreover, the impact of the amplifier noise on the phase PSDs agrees well with the numerical simulations shown in Fig.~\ref{fig:FNvsPin}. 

For the phase PSD obtained using the EKF phase estimation method, the impact of amplifier noise is observable for the input power levels ranging from -25 to -15 dBm. However, we would like to stress that the EKF is able to filter out a significant amount of amplifier noise affecting the phase, thus resulting in PSDs that resemble the reference to a high degree. This is also in accordance with the numerical results. For the input signal power of -10 dBm or higher, we did not observe any difference compared to the reference PSD. We have therefore avoided to plot it.  

Finally, in Fig.~\ref{fig:spec broad vs Pin exp}, we compute the corresponding spectral broadening, induced by the EDFA, with respect to the signal after the MOPA, and plot it as a function of the input signal power to the EDFA. The spectral broadening is computed by the numerical integration of the phase PSD and is denoted by $\Delta \nu^{\tan^{-1},int.}_a$ and $\Delta \nu^{EKF,int.}_a$. For the reference, we also plot theoretically minimum achievable spectral broadening, i.e.~$\Delta \nu_a$ and $\Delta \nu^{MAP}_a$.   

It is worth noting that the computed spectral broadening $\Delta \nu^{\tan^{-1},int.}_a$ closely follows the theoretical limit $\Delta \nu_a$. It is also observed that the spectral broadening obtained using the EKF phase estimation method, $\Delta \nu^{EKF,int.}_a$, is significantly lower compared to $\Delta \nu^{\tan^{-1},int.}_a$. This is in accordance with the numerical results. By employing the EKF based phase estimation method a reduction of 44 dB in terms of the spectral broadening is achieved, as illustrated in Fig.~\ref{fig:spec broad vs Pin exp}. Finally, $\Delta \nu^{EKF,int.}_a$ exhibits a penalty of approximately 15 dB compared to $\Delta \nu^{MAP}_a$. The penalty may be attributed to the fact that the state-space model employed for the phase estimation assumes that the laser is only dominated by the intrinsic noise sources.
%%%%%%%%%%%%%%%%%%%%%%%%%%%%%%%%%%%%%%%%%%%%%%%%%%%%%%%%%% section 5 %%%%%%%%%%%%%%%%%%%%%%%%%%%%%%%%%%%%%%%%%%%%%%%%%%%%%%%%%%
\section{Conclusion} \label{sec:conc}
The impact of amplifier noise on the phase fluctuation, and the corresponding spectral broadening, of the incoming signal has been investigated theoretically, numerically and experimentally. We have shown that a heterodyne signal detection in combination with the extended Kalman filter is a realization of a practical optimum phase measurement method in the presence of amplifier noise. The proposed phase measurement method has a relatively small penalty compared to the quantum limit. Most importantly, a significant reduction of the impact of the amplifier noise has been demonstrated, both numerically and experimentally, compared to when using a widely deployed (sub-optimal) phase estimation method. 

\section{Future work} \label{sec:future}
The analysis presented in the paper is only valid for coherent signals which implies that the input to the amplifier is provided by a coherent laser source (coherent-state light). For nonclassical states, the analysis needs to be revised which indeed is a very interesting follow-up subject.

Moreover, for future work, it would be highly beneficial to consider the real-time implementation of the proposed EKF phase estimation method. Such an implementation should be feasible as the state-space model for the EKF is relatively simple (single state and measurement equation). However, to implement it at high frequencies (>10 GHz) a high degree of parallelization would be needed which may lead to an increased penalty compared to the quantum limit.

\medskip
\noindent\textbf{Funding.} This work was supported by the European Research Council (ERC CoG FRECOM grant 771878), the Villum Foundations (VYI OPTIC-AI grant no. 29344).

%\section*{Acknowledgments}
%The authors thank H. Haase, C. Wiede, and J. Gabler for technical support.

\medskip
\noindent\textbf{Disclosures.} The authors declare no conflicts of interest.

%\section*{Supplemental Documents}
%\emph{Optica} authors may include supplemental documents with the primary manuscript. For details, see \href{http://www.opticsinfobase.org/submit/style/supplementary-materials-optica.cfm}{Supplementary Materials in Optica}. To reference the supplementary document, the statement ``See Supplement 1 for supporting content.'' should appear at the bottom of the manuscript (above the references).

\bigskip \noindent See \href{link}{Supplement 1} for supporting content.

% Bibliography
\bibliography{references}

\end{document}